\documentclass{PoS}
\usepackage{cite}
\usepackage[english]{babel}
\usepackage{amsmath}
\usepackage[cal=cmcal,greekuppercase=italicized]{mathdesign}
\usepackage{slashed}
\usepackage{dsfont}
\usepackage{graphicx}
\usepackage{xcolor}

\makeatletter
\DeclareRobustCommand*{\bfseries}{%
  \not@math@alphabet\bfseries\mathbf
  \fontseries\bfdefault\selectfont
  \boldmath
}
\makeatother

\newcommand{\Ll}{\ensuremath{\mathrm{L}}}
\newcommand{\Rr}{\ensuremath{\mathrm{R}}}

\newcommand{\dq}{\ensuremath{\mathrm{d}}}
\newcommand{\uq}{\ensuremath{\mathrm{u}}}
\newcommand{\el}{\ensuremath{\mathrm{e}}}
\newcommand{\nul}{\ensuremath{\mathrm{\nu}}}

\newcommand{\im}{\ensuremath{\mathrm{i}\,}}

\newcommand{\diag}{\operatorname{diag}}
\newcommand{\SU}{\ensuremath{\mathrm{SU}}}
\newcommand{\U}{\ensuremath{\mathrm{U}}}

\newcommand{\eV}{\ensuremath{\mathrm{eV}}}
\newcommand{\GeV}{\ensuremath{\mathrm{GeV}}}

\newcommand{\hc}{\ensuremath{\text{h.\,c.\ }}}
\newcommand{\tp}{\ensuremath{\text{T}}}

\newcommand{\Lag}{\ensuremath{\mathcal{L}}}

\newcommand{\ie}{i.\,e.~}

\title{Neutrino Mixing from SUSY breaking\footnote{Talk given at the Summer
    School and Workshop on the Standard Model and Beyond 2013.}}

\ShortTitle{Neutrino Mixing from SUSY breaking}

\author{\speaker{Wolfgang~G.~Hollik}\\
  Institut f\"ur Theoretische Teilchenphysik\thanks{Report number: TTP15-016}\\
  Karlsruhe Institute of Technology\\
  E-mail: \email{wolfgang.hollik@kit.edu}}

\abstract{We propose a mechanism to generate the neutrino mixing matrix
  from supersymmetric threshold corrections. Flavor violating soft
  breaking terms induce flavor changing self-energies that give a finite
  renormalization to the mixing matrix. The described threshold
  corrections get enhanced in case of quasi-degenerate neutrino
  masses. In this scenario, we adjust potentially arbitrary
  soft breaking parameters in a way to reproduce the observed neutrino
  mixing at one loop working with non-minimal flavor violating soft
  parameters. To incorporate small neutrino masses already at tree-level
  via a type I seesaw mechanism, we extend the Minimal Supersymmetric
  Standard Model with singlet Majorana neutrinos. The radiative
  corrections do not decouple with the scale of Supersymmetry and
  persist when the spectrum is shifted to higher values. Moreover, the
  mixing matrix renormalization with flavor-changing self-energies is
  not restricted to supersymmetric theories and give similar results in
  any theory with new flavor structures.}

\FullConference{Proceedings of the Corfu Summer Institute 2014 "School
  and Workshops on Elementary Particle Physics and Gravity",\\
  3-21 September 2014\\
  Corfu, Greece }

\begin{document}

\section{Introduction}
The Standard Model (SM) of elementary particle physics provides excellent
predictions of fundamental properties of nature and elementary
processes. Most parameters of the theory are related to flavor and have
to be determined by measurement: fermion masses and mixing angles. In
general, they are arbitrary parameters of the theory originating in the
Yukawa couplings of the fermions to the SM Higgs doublet. Gauge
interactions do not change the flavor, therefore the (weak) gauge
interaction basis sets the flavor basis. Masses are generated after
spontaneous symmetry breaking through the Yukawa couplings
\begin{equation}\label{eq:SMYuk}
- \Lag^\text{SM}_\text{Yuk} = Y^\dq_{ij} \bar Q_{\Ll, i} \cdot H d_{\Rr,
  j} - Y^\uq_{ij} \bar Q_{\Ll, i} \cdot \tilde H u_{\Rr, j} + Y^\el_{ij}
\cdot \bar L_{\Ll, i} \cdot H e_{\Rr, j} - Y^\nul_{ij} \bar L_{\Ll, i}
\cdot \tilde H \nu_{\Rr, j} + \hc,
\end{equation}
where \(H = (h^+, h^0)^\tp\) is the Higgs doublet and \(\tilde H = \im
\tau_2 H^*\) the charge conjugated version of it. The quark and lepton
left-handed doublets are given by \(Q_\Ll = (u_\Ll, d_\Ll)^\tp\) and
\(L_\Ll = (\nu_\Ll, e_\Ll)^\tp\), respectively. The right-handed SM
fermions are labeled obviously. For later purpose we have already
introduced right-handed neutrinos \(\nu_\Rr\) and their Yukawa couplings
to left-handed leptons. Right-handed neutrinos are complete singlets
under the SM gauge group. Fermion masses are given by the mass matrices
\(m^f = v Y^f / \sqrt{2}\) with the vacuum expectation value \(v\) of
the Higgs field, \(\langle h^0\rangle = v/\sqrt{2} = 174\,\GeV\) and \(f
= \uq, \dq, \el, \nul\). Generation indices \(i,j\) count the number of
generations and diagonalization of the mass matrices \(m^f\) transform
into the mass eigenbasis. We can do so with bi-unitary transformations,
such that
\begin{equation}
Y^f \to S_\Ll^f Y^f \left( S_\Rr^f \right)^\dag = \hat Y^f = \text{diagonal}.
\end{equation}
We find the fermion mixing matrices as they appear in the weak
charged current due to misalignment of the Yukawa couplings as the
Cabibbo--Kobayashi--Maskawa \cite{Cabibbo:1963yz, Kobayashi:1973fv}
(CKM) matrix \(V_\text{CKM} = S^\uq_\Ll \left(S^\dq_\Ll\right)^\dag\)
and the Pontecorvo--Maki--Nakagawa--Sakata \cite{Pontecorvo:1957cp,
  Maki:1962mu} (PMNS) matrix \(U_\text{PMNS} = S^\el_\Ll
\left(S^\nul_\Ll\right)^\dag\).

The observed mixing patterns for quarks and leptons are quite
different. Where the CKM matrix is close to the unit matrix consistent
with small mixing, the PMNS matrix shows a rather anarchic mixing
pattern with no clear hierarchy in the elements. This unlike behavior of
quark versus lepton mixing can be displayed very visually showing the
sizes of the magnitudes
\[
\left|V_\text{CKM}\right| = \left(\begin{array}{ccc}
        \begin{picture}(10,10)\put(5,5){\circle*{9.743}}\end{picture}
        & \begin{picture}(10,10)\put(5,5){\circle*{2.254}}\end{picture}
        & \begin{picture}(10,10)\put(5,5){\circle*{.0355}}\end{picture} \\
        \begin{picture}(10,10)\put(5,5){\circle*{2.252}}\end{picture}
        & \begin{picture}(10,10)\put(5,5){\circle*{9.734}}\end{picture}
        & \begin{picture}(10,10)\put(5,5){\circle*{.414}}\end{picture} \\
        \begin{picture}(10,10)\put(5,5){\circle*{.0886}}\end{picture}
        & \begin{picture}(10,10)\put(5,5){\circle*{.0405}}\end{picture}
        & \begin{picture}(10,10)\put(5,5){\circle*{9.991}}\end{picture}
\end{array}\right)\;,
\qquad
\left|U_\text{PMNS}\right| = \left(\begin{array}{ccc}
        \begin{picture}(10,10)\put(5,5){\circle*{8.40}}\end{picture}
        & \begin{picture}(10,10)\put(5,5){\circle*{5.22}}\end{picture}
        & \begin{picture}(10,10)\put(5,5){\circle*{1.47}}\end{picture} \\
        \begin{picture}(10,10)\put(5,5){\circle*{4.88}}\end{picture}
        & \begin{picture}(10,10)\put(5,5){\circle*{6.09}}\end{picture}
        & \begin{picture}(10,10)\put(5,5){\circle*{6.26}}\end{picture} \\
        \begin{picture}(10,10)\put(5,5){\circle*{2.37}}\end{picture}
        & \begin{picture}(10,10)\put(5,5){\circle*{5.97}}\end{picture}
        & \begin{picture}(10,10)\put(5,5){\circle*{7.66}}\end{picture}
\end{array}\right).
\]

The structure of the CKM matrix suggests the possibility of generating
quark mixing via higher orders in perturbation theory: the flavor
violating contributions are small and of the size of typical one-loop
corrections. The PMNS matrix on the other side does not allow for lepton
mixing as a genuine loop effect at first sight: the leptonic mixing
angles are just too large. However, the contributions to the mixing
matrix renormalization can be drastically enhanced by the neutrino mass
spectrum as will be discussed in the following after a brief overview of
a radiative description of quark mixing. We follow the procedure for the
supersymmetric renormalization of the CKM matrix \cite{Crivellin:2008mq}
and apply it to the lepton case where the contributions are generically
smaller because only slepton--electroweakino loops are present. As we
will show, the neutrino corrections can be significantly enhanced and
therefore dominating over any tree-level mixing pattern.

\section{Radiative Flavor Violation in the MSSM}\label{sec:RFV}
A radiative origin of quark masses and flavor mixing was already
proposed in the early days of the SM~\cite{Weinberg:1972ws} and later on
applied in the context of grand unified theories~\cite{Segre:1981nj,
  Ibanez:1982xg, Ibanez:1981nw}. Especially supersymmetric models give
the opportunity to radiatively generate masses and Yukawa couplings via
soft breaking terms~\cite{Lahanas:1982et, Masiero:1983ph, Banks:1987iu,
  Ma:1988fp, Borzumati:1999sp}. The soft breaking Lagrangian of
supersymmetric theories generically carry arbitrary flavor structures
in addition to the flavor structure of the SM which is transferred from
Eq.~\eqref{eq:SMYuk} to the \emph{superpotential} of the Minimal
Supersymmetric Standard Model (MSSM):
\begin{equation}
  \mathcal{W}_\text{MSSM} = \mu H_\dq \cdot H_\uq
  - Y^\dq_{ij} H_\dq Q_{\Ll, i} \bar D_{\Rr, j}
  + Y^\uq_{ij} H_\uq \cdot Q_{\Ll, i} \bar U_{\Rr, j}
  - Y^\el_{ij} H_\dq \cdot L_{\Ll, i} \bar E_{\Rr, j}
  + Y^\nul_{ij} H_\uq \cdot L_{\Ll, i} \bar N_{\Rr, j},
\end{equation}
where the fields are chiral superfields and the number of Higgs doublets
is doubled in the MSSM. Charge conjugated right-handed matter fermions
fit into the left-chiral superfields \(\bar U_\Rr = \{\tilde u_\Rr^*,
u_\Rr^c\}\), \(\bar D_\Rr = \{\tilde d_\Rr^*, d_\Rr^c\}\), \(\bar E_\Rr
= \{\tilde e_\Rr^*, e_\Rr^c\}\) and \(\bar N_\Rr = \{\tilde\nu_\Rr^*,
\nu_\Rr^c\}\), whereas \(Q_\Ll = \{\tilde q_\Ll, q_\Ll\}\) and \(L_\Ll =
\{\tilde\ell_\Ll, \ell_\Ll\}\) represent the weak doublets. The Higgs
doublets \(H_\uq = (H_\uq^+, H_\uq^0)^\tp\) and \(H_\dq = (H_\dq^0,
H_\dq^-)^\tp\) give masses to the up-type and down-type fermions,
respectively.

Supersymmetry (SUSY) is softly broken via the following soft breaking
terms
\begin{equation}\label{eq:softMSSM}
\begin{aligned}
  -\Lag^\text{MSSM}_\text{soft} =&~~ \tilde q_{\Ll,i}^*
  \left(\tilde{m}^2_Q\right)_{ij} \tilde q_{\Ll,j} + \tilde u_{\Rr,i}^*
    \left(\tilde{m}^2_u\right)_{ij} \tilde u_{\Rr,j}
    + \tilde d_{\Rr,i}^* \left(\tilde{m}^2_d\right)_{ij} \tilde d_{\Rr,j} \\
    & + \tilde \ell_{\Ll,i}^* \left(\tilde{m}^2_\ell\right)_{ij}
    \tilde\ell_{\Ll,j} + \tilde e_{\Rr,i}^*
    \left(\tilde{m}^2_e\right)_{ij} \tilde e_{\Rr,j} + \tilde
    \nu_{\Rr,i}^* \left(\tilde{m}^2_\nu\right)_{ij} \tilde
    \nu_{\Rr,j} \\
    & + \bigg[ h_\dq \cdot \tilde \ell_{\Ll,i} A^\el_{ij} \tilde
    e^*_{\Rr,j} + \tilde \ell_{\Ll,i} \cdot h_\uq A^\nul_{ij} \tilde
    \nu^*_{\Rr,j} + h_\dq \cdot \tilde q_{\Ll,i} A^\dq_{ij} \tilde
    d^*_{\Rr,j} + \tilde q_{\Ll,i} \cdot h_\uq A^\uq_{ij} \tilde
    u^*_{\Rr,j} + \hc \bigg ].
\end{aligned}
\end{equation}
We omitted soft breaking terms relevant for gauginos and Higgses. The
soft breaking masses \(\tilde{m}_f^2\) as well as the trilinear
couplings \(A^f\) have \emph{a priori} no restriction in their flavor
structure which causes problems in flavor changing neutral current
observables. So either one imposes Minimal Flavor Violation (MFV) which
reduces dangerously large contributions to those observables under the
assumption that only SM Yukawa couplings transport flavor information.
In that view, the trilinear couplings are \(A^f = a_f Y^f\) with some
SUSY-scale parameter \(a_f\) that is flavor-universal. On the other
hand, we can seek for a symmetry that forbids flavor transitions in the
SM sector at tree-level and generates all flavor violation radiatively
(RFV). In that view, all the SM Yukawa couplings are simultaneously
diagonal. Such symmetries may be \(\U(2)\) flavor symmetries as proposed
and studied in~\cite{Ferrandis:2004ng, Ferrandis:2004ri,
  Crivellin:2008mq, Crivellin:2009pa, Crivellin:2011sj,
  Crivellin:2011fb, Altmannshofer:2014qha}. The Yukawa couplings vanish
except for the third generation
\begin{equation}
Y^f = \begin{pmatrix} 0 & 0 & 0 \\ 0 & 0 & 0 \\ 0 & 0 & y_f \end{pmatrix},
\end{equation}
and the first two generation Yukawa couplings and the mixing have to be
generated radiatively.

\begin{figure}[tb]
\begin{minipage}{.5\textwidth}
\includegraphics[width=\textwidth]{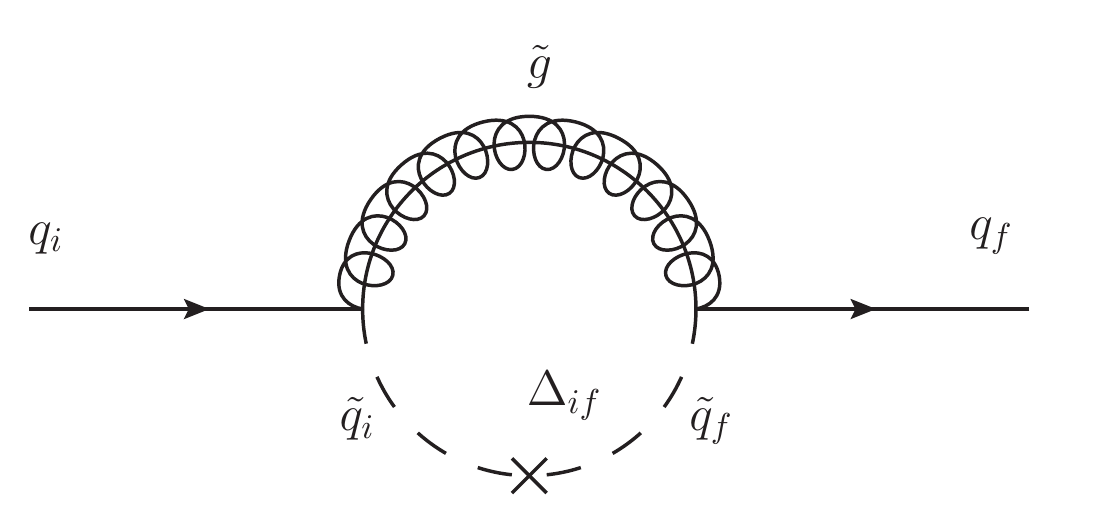}
\end{minipage}%
\begin{minipage}{.5\textwidth}
\includegraphics[width=\textwidth]{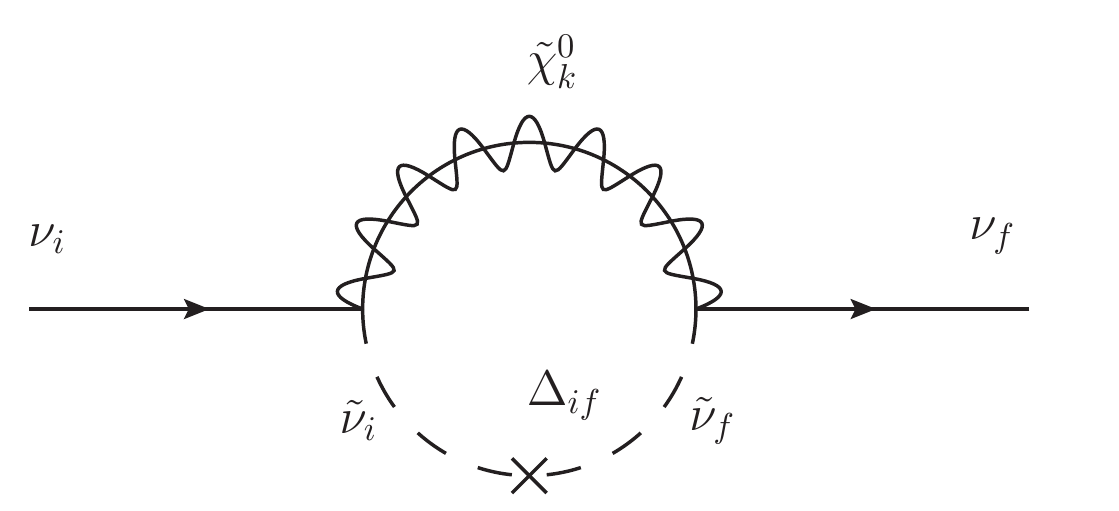}
\end{minipage}%
\caption{Flavor changing self-energies for quarks (left, squark--gluino
  loop) and neutrinos (right, slepton--gaugino/higgsino loop).}
\label{fig:selfen}
\end{figure}

We want to focus on the supersymmetric renormalization of the CKM matrix
according to~\cite{Crivellin:2008mq} and leave away the radiative
generation of Yukawa couplings. By virtue of the SUSY one-loop
corrections, the flavor transitions enter via flavor changing
self-energies \(\Sigma_{fi}^\psi\) for the fermion \(\psi\) as shown in
Fig.~\ref{fig:selfen}. We have drawn the quark self-energy in the flavor
basis, where the quark--squark--gluino interaction is diagonal---the
flavor change sits in the mass insertion \(\Delta^{\tilde q}_{if}\)
related to the squark mass matrix. A similar diagram can be drawn for
the bino/wino-like neutralino and also exists for quarks, which is
subdominant there, but the only contributing self-energy for neutrinos
(including the other neutralinos and chargino--slepton loops). The
fermion self-energies can be decomposed as
\begin{equation}
  \Sigma^\psi_{fi} (p) = \Sigma_{fi}^{\psi, RL} (p^2)\; P_\Ll +
  \Sigma_{fi}^{\psi, LR} (p^2)\; P_\Rr + \slashed{p} \left[
    \Sigma_{fi}^{\psi, LL} (p^2)\; P_\Ll + \Sigma_{fi}^{\psi, RR} (p^2)\; P_\Rr
  \right],
\label{eq:SelfEnDec}
\end{equation}
with the left- and right-handed projectors
\(P_{\Ll,\Rr}\). Renormalization of the mixing matrix follows the
prescription of~\cite{Denner:1990yz} as done
in~\cite{Crivellin:2008mq}. The renormalized CKM matrix is found to be
\begin{equation}
V = \left( \mathds{1} + {\Delta U^u_L}^\dag \right) V^{(0)} \left(
  \mathds{1} + \Delta U^d_L \right),
\end{equation}
where \(V^{(0)}\) is the unrenormalized, ``bare'' mixing matrix (the
quark mixing matrix at tree-level) and the \(\Delta U^q_L\) are defined
via their contribution to the weak charged current vertex
\[
\im \frac{g_2}{\sqrt{2}} \gamma^\mu P_\Ll V^{(0)} \quad \to \quad
\im \frac{g_2}{\sqrt{2}} \gamma^\mu P_\Ll \left( V^{(0)} + D_L + D_R \right)
\]
with
\begin{equation}
D_{L,fi} = \sum_{j=1}^3 V^{(0)}_{fj} \left[ \Delta U^d_L \right]_{ji}
\qquad \text{and} \qquad
D_{R,fi} = \sum_{j=1}^3 \left[ {\Delta U^u_L}^\dag \right]_{fj} V^{(0)}_{ji}.
\end{equation}
The contributions \(D_{L,R}\) are calculated in terms of the components
in Eq.~\eqref{eq:SelfEnDec}
\begin{equation}\label{eq:DLdown}
D_{L,fi} = \sum_{j\neq i} V^{(0)} \frac{ m_{d_j} \left( \Sigma^{d,
      RL}_{ji} + m_{d_i} \Sigma^{d, RR}_{ji} \right) + m_{d_i} \left(
    \Sigma^{d, LR}_{ji} + m_{d_i} \Sigma^{d, LL} \right)}{m_{d_i}^2 - m_{d_j}^2}.
\end{equation}
A similar expression holds for \(D_{R,fi}\) with \(d \to u\). The large
hierarchy in quark masses allows to expand in small ratios as \(m_s /
m_b\), in general \(m_{q_i} / m_{q_j}\) with \(i<j\), and one has
(neglecting \(\Sigma^{q, LL/RR}\) which are relatively suppressed by the
heavy masses in the loop, \ie the overall SUSY mass scale)
\begin{equation}\label{eq:DeltaUq}
\Delta U^q_L = \begin{pmatrix}
0 & \frac{1}{m_{q_2}} \Sigma^{q, LR}_{12} & \frac{1}{m_{q_3}} \Sigma^{q, LR}_{13} \\
\frac{-1}{m_{q_2}} \Sigma^{q, RL}_{21} & 0 & \frac{1}{m_{q_3}} \Sigma^{q, LR}_{23} \\
\frac{-1}{m_{q_3}} \Sigma^{q, RL}_{31} & \frac{-1}{m_{q_3}} \Sigma^{q, RL}_{32} & 0
\end{pmatrix}.
\end{equation}
This is not true in the neutrino case as we shall see in the following section.

\section{Radiative Lepton Flavor Violation in the \(\mathbf{\nul}\)MSSM}
The description of Radiative Lepton Flavor Violation (RLFV) follows
basically the setup of Sec.~\ref{sec:RFV}. If the neutrino sector were
just a copy of the up-type quark sector, we would be done and could
discuss the phenomenological output of RLFV. However, things are likely to
be different.

Though nothing is odd with mirroring of what we have reviewed to the
neutrino sector, there is one puzzle connected to neutrinos: in the SM
they are exactly massless. Experiment nevertheless tells us that they at
least have tiny masses~\cite[and references
therein]{Agashe:2014kda}. This can be accomplished on the one hand by
setting the neutrino Yukawa couplings to a small value. On the other
hand, this smells artificial. A tree-level solution to that puzzle was
proposed via an effective operator~\cite{Weinberg:1979sa}
\begin{equation}\label{eq:WeinbergOp}
\Lag_\text{dim 5} = \frac{\lambda_{ij}}{\Lambda} \left( L_i \cdot H
\right) C \left( H \cdot L_j \right),
\end{equation}
where the couplings \(\lambda_{ij}\) are dimensionless but
\(\mathcal{O}(1)\) couplings. Neutrino masses are generated after
spontaneous symmetry breaking as \(m_\nu = v^2 \lambda / (2 \Lambda)\)
and are suppressed by the scale \(\Lambda\) which can be much larger
than the electroweak scale, \(\Lambda \gg v\). The matrix \(C\) in
Eq.~\eqref{eq:WeinbergOp} is the charge conjugation matrix: neutrino
masses generated via this mechanism are Majorana masses. UV complete
models leading to Eq.~\eqref{eq:WeinbergOp} have been elaborated with
additional fermions and scalars~\cite{Minkowski:1977sc, Mohapatra:1979ia,
  Magg:1980ut, Schechter:1980gr, Schechter:1981cv, Foot:1988aq}. We
shall extend the MSSM with three singlet chiral superfields that act as
right-handed neutrinos and get a Majorana mass term in the
superpotential
\begin{equation}
\mathcal{W}_{\nul\text{MSSM}} = \mathcal{W}_\text{MSSM} + \frac{1}{2}
M^\Rr_{ij} \bar N_{\Rr, i} \bar N_{\Rr, j}.
\end{equation}
We refer to this model as \(\nul\)MSSM. The effective neutrino mass
matrix is given by the combination
\begin{equation}
m_\nul = -\frac{v^2}{2}\; Y_\nul\; M_\Rr^{-1}\; Y^\tp_\nul.
\end{equation}
Without loss of generality, we can choose both the charged lepton Yukawa
couplings \(Y_\el\) and the right-handed Majorana mass matrix \(M_\Rr\)
diagonal. The weak mixing matrix is then determined by the
diagonalization of \(m_\nul\):
\begin{equation}
\hat m_\nul = U^*_\text{PMNS}\; m_\nul\; U^\dag_\text{PMNS} = \text{diagonal}.
\end{equation}
We do not know exactly the masses of the light neutrinos. However, out
of neutrino oscillations, mass squared differences \(\Delta m_{ij}^2 =
m_{\nu_i}^2 - m_{\nu_j}^2\) can be obtained and therewith in principle
the spectrum calculated, here in case of a ``normal'' ordering
(``inverted'' ordering has \(|m_{\nu_3}| = m_\nul^{(0)}\)):
\begin{equation}
\left|m_{\nu_1}\right| = m_\nul^{(0)}, \quad
\left|m_{\nu_2}\right| = \sqrt{\big(m_\nul^{(0)}\big)^2 + \Delta m_{21}^2}, \quad
\left|m_{\nu_3}\right| = \sqrt{\big(m_\nul^{(0)}\big)^2 + \Delta m_{31}^2}.
\end{equation}
What we do not know is the mass of the lightest neutrino
\(m_\nul^{(0)}\). The \(\Delta m^2_{ij}\) are known and
\begin{equation}\label{eq:Deltam2}
\Delta m_{21}^2 = 7.50^{+0.19}_{-0.17} \times 10^{-5}\,\eV^2, \qquad
\Delta m_{31}^2 = 2.457\pm 0.047 \times 10^{-3}\,\eV^2,
\end{equation}
as follows from a global fit on neutrino
data~\cite{Gonzalez-Garcia:2014bfa}. We keep \(m_\nul^{(0)}\) as free
parameter.

\begin{figure}[tb]
\includegraphics[width=\textwidth]{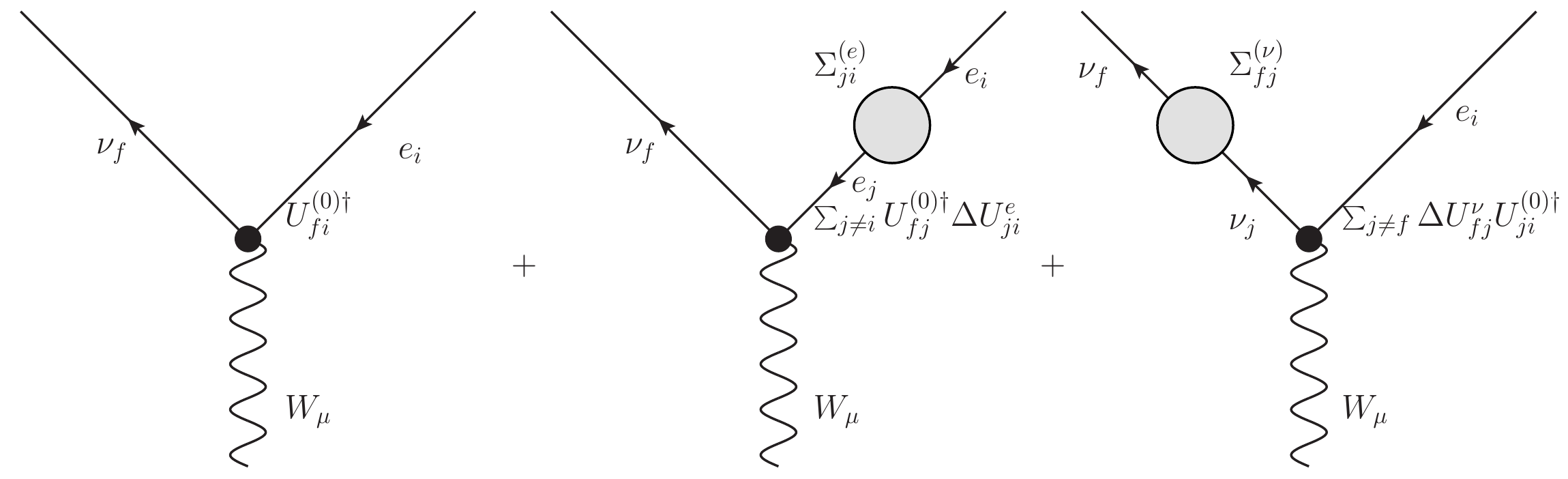}
\caption{Renormalization of the mixing matrix by means of flavor
changing self-energies at external legs according to
\cite{Denner:1990yz}. Shown are diagrams for the lepton mixing matrix.
The inclusion of finite mixing matrix counterterms \(\Delta
U^{\nu,e}_{ij}\) may significantly alter the tree-level mixing matrix
\(U^{(0)}_{ij}\). Similar diagrams with obvious changes in the labels can
be obtained for quarks.}\label{fig:PMNSreno}
\end{figure}

Before we calculate the SUSY one-loop contribution to the mixing matrix
renormalization, we have to keep in mind, that our neutrinos are
Majorana fermions which is a property that relates the components of
Eq.~\eqref{eq:SelfEnDec} in a way that \(\Sigma^\nul_{LR} =
{\Sigma^\nul_{RL}}^*\) and \(\Sigma^\nul_{RR} =
{\Sigma^\nul_{LL}}^*\). Majorana masses and self-energies are symmetric,
so \(\Sigma^\nul_{fi} = \Sigma^\nul_{if}\). We decompose the neutrino
self-energy in a ``scalar'' (\(S\)) and ``vectorial'' (\(V\)) part
\begin{equation}\label{eq:SigmaNu}
  \Sigma^{\nul}_{fi} (p) =
  \Sigma^{\nul, S}_{fi} (p^2)\; P_\Ll + {\Sigma^{\nul, S}_{fi}}^* (p^2)\; P_\Rr
  + \slashed{p} \left[
    \Sigma^{\nul, V}_{fi} (p^2)\; P_\Ll + {\Sigma^{\nul, V}_{fi}}^* (p^2)\; P_\Rr
\right].
\end{equation}
Since in the definition of the lepton mixing matrix up and down are
interchanged, we have
\begin{equation}\label{eq:PMNSreno}
  U_\text{PMNS} = \left(\mathds{1} + \Delta U_L^\el\right)
  U^{(0)} \left(\mathds{1} + \Delta U_L^\nul\right)^\dag
  \approx\; U^{(0)} + \left(\Delta U^\el_L\right)
  U^{(0)} + U^{(0)} \left(\Delta U^\nul_L\right)^\dag,
\end{equation}
reflecting the decomposition of the mixing matrix in contributions from
charged and neutral leptons, \(U_\text{PMNS} = S^\el_\Ll
\left(S^\nul_\Ll\right)^\dag\). The contribution from the neutrino leg
(the ``left'' leg in Fig.~\ref{fig:PMNSreno}) is given analogously to
Eq.~\eqref{eq:DLdown} by
\begin{equation}\label{eq:PMNSrenNU}
  D_{L,fi} = \sum_{j=1}^n \left[\Delta U_L^{\nul}\right]_{fj} U^{(0)\dag}_{ji}
  = \sum_{j \neq f} \frac{m_{\nul_f} \left( \Sigma^{\nul, S}_{fj}
      + m_{\nul_f} \Sigma^{\nul, V}_{fj} \right) + m_{\nul_j} \left(
      {\Sigma^{\nul, S}_{fj}}^* + m_{\nul_f} {\Sigma^{\nul, V}_{fj}}^*
    \right)}{ m_{\nul_j}^2-m_{\nul_f}^2} U^{(0)\dag}_{ji}.
\end{equation}
Neglecting \(\Sigma^{\nu, V}\) (which again is relatively suppressed
with \(1/M_\text{SUSY}\)), we have
\begin{equation}
\left[\Delta U_L^{\nul}\right]_{fi} =
  \frac{m_{\nul_f} \Sigma^{\nul, S}_{fi}
    + m_{\nul_i} {\Sigma^{\nul, S}_{fi}}^*}{m_{\nul_i}^2-m_{\nul_f}^2}.
\end{equation}
In view of this result, we find an enhancement of the contribution
\(\Delta U^\nul_L\) in case of quasi-degenerate neutrino
masses~\cite{Hollik:2014lka}. The neutrino self-energies \(\Sigma^{\nu,
  S}\) are of the same order as the neutrino masses, so \(\left[ \Delta
    U^\nul_L \right]_{fi} \sim m_{\nul_f} m_{\nul_i} / \Delta m^2_{fi}\)
  which can be as large as \(5 \times 10^3\) for \(m_\nul^{(0)} =
  0.35\,\eV\) and \(f,i = 1,2\).

The charged lepton contribution, in contrast, obeys a hierarchical
structure like Eq.~\eqref{eq:DeltaUq} which give at most CKM-like mixing
and can be neglected in the further discussion while the neutrino-leg
contribution is the dominating one in case of quasi-degenerate
neutrinos. For hierarchical neutrino masses, flavor changing threshold
corrections to the mixing matrix as described above are sub-dominant.
However, hierarchical masses may ask for a special treatment which can
be combined with the radiative Yukawa couplings discussed in the
introduction: a successive breaking of an initial \(\left[ \U(3)
\right]^6\) flavor symmetry leads to a description of fermion mixing in
terms of mass ratios only, that can be related to the symmetry breaking
parameters~\cite{Hollik:2014jda}.

We do not give analytic expressions for self-energies in this
article since such results can be found in \cite{Dedes:2007ef} and
\cite{Hollik:2014hya}. Instead, we shall discuss the flavor structure of
the contributing one-loop diagrams and show that we observe a
\emph{non-decoupling} effect with respect to the SUSY scale: if all
SUSY-scale parameters are uniformly shifted to higher values, the
results do not change.

\begin{figure}[tb]
\includegraphics[width=\textwidth]{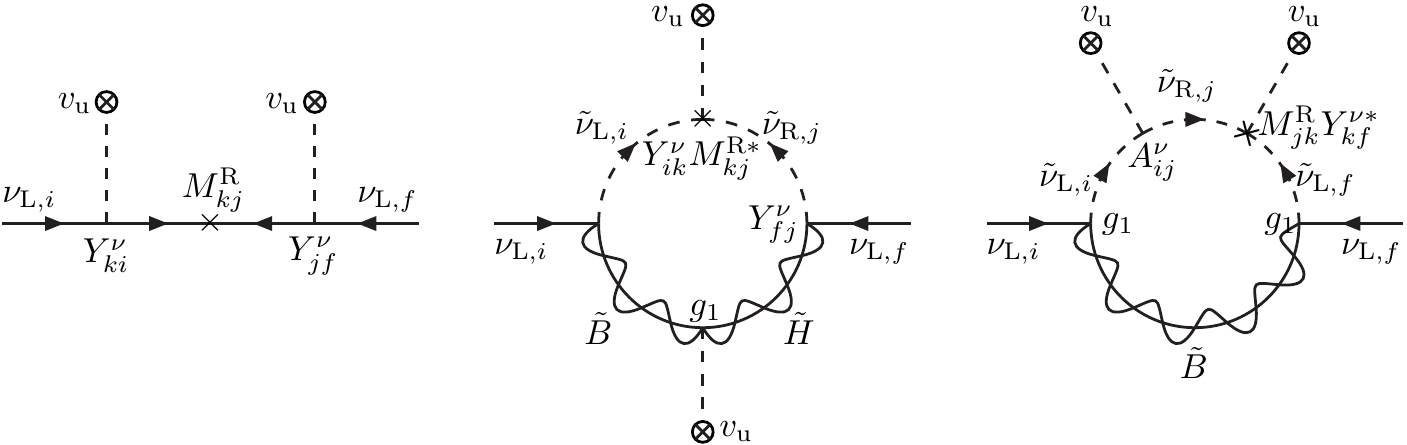}
\caption{Tree-level and one-loop contributions to the seesaw neutrino
  mass. Insertions of the right-handed Majorana mass \(M_\Rr\) provide
  the lepton number violating fermion flip.}\label{fig:seesawreno}
\end{figure}

Diagrams contributing to the renormalization of the PMNS matrix and the
one-loop neutrino masses are given in Fig.~\ref{fig:seesawreno}, where
the first one shows the tree-level contributions and the second and
third diagram (from left to right) two phenomenologically different
one-loop self-energies. The second diagram leaves the mixing matrix
invariant in case of degenerate right-handed neutrinos as can be seen
from the structure
\begin{equation}\label{eq:mnulog}
m_\nul^\text{tree + 1-loop} = v_\uq^2 Y_\nul \diag\left(
  \frac{1}{M_{\Rr, k}} + \frac{g_1^2}{64\pi^2} \frac{\log\left(
      \frac{M_\text{SUSY}^2}{M^2_{\Rr, k}} \right)}{M_{\Rr, k}} \right) Y_\nul^\tp,
\end{equation}
in the limit of degenerate SUSY masses and with the approximation
\(m_{\tilde \nu_{\Rr, k}} = M_{\Rr, k}\) where the difference is of
\(\mathcal{O}(M_\text{SUSY})\). With degenerate right-handed masses, the
diagonal matrix in Eq.~\eqref{eq:mnulog} is proportional to the unit
matrix and therefore does not alter the diagonalization of \(Y_\nul
Y_\nul^\tp\).

\begin{figure}[tb]
\includegraphics[width=\textwidth]{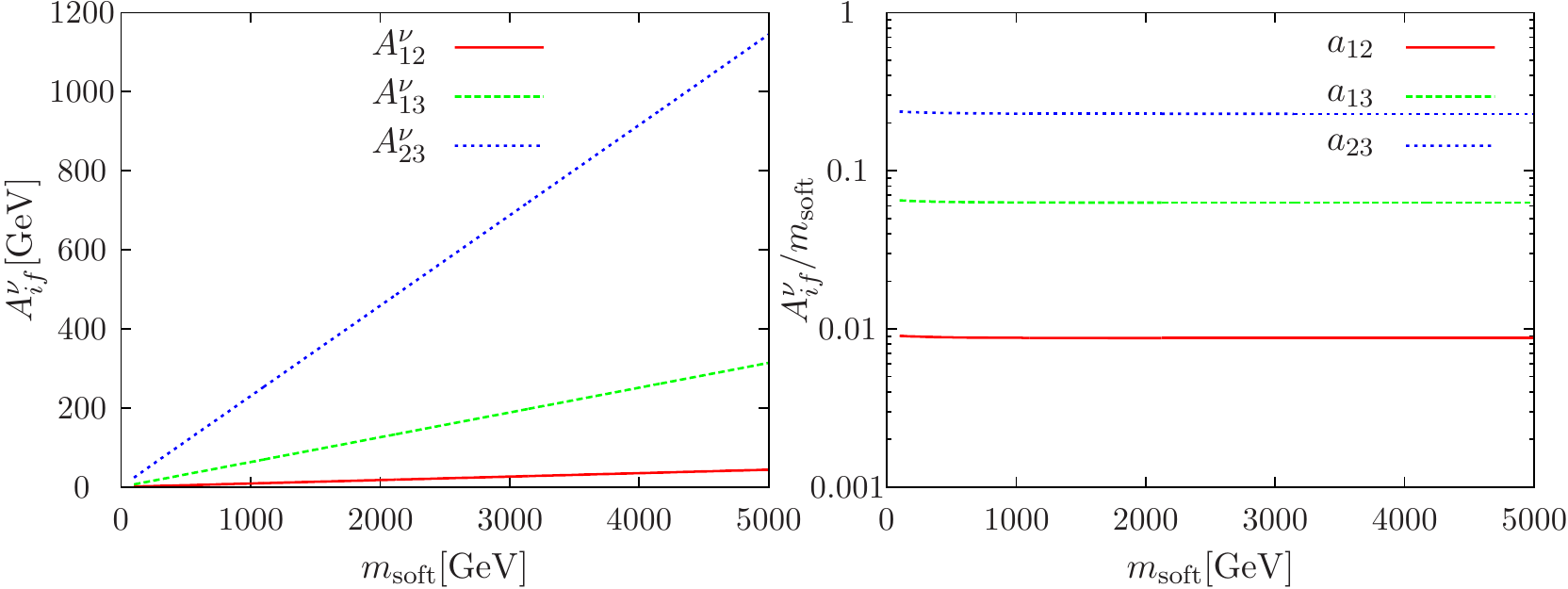}\\
\includegraphics[width=\textwidth]{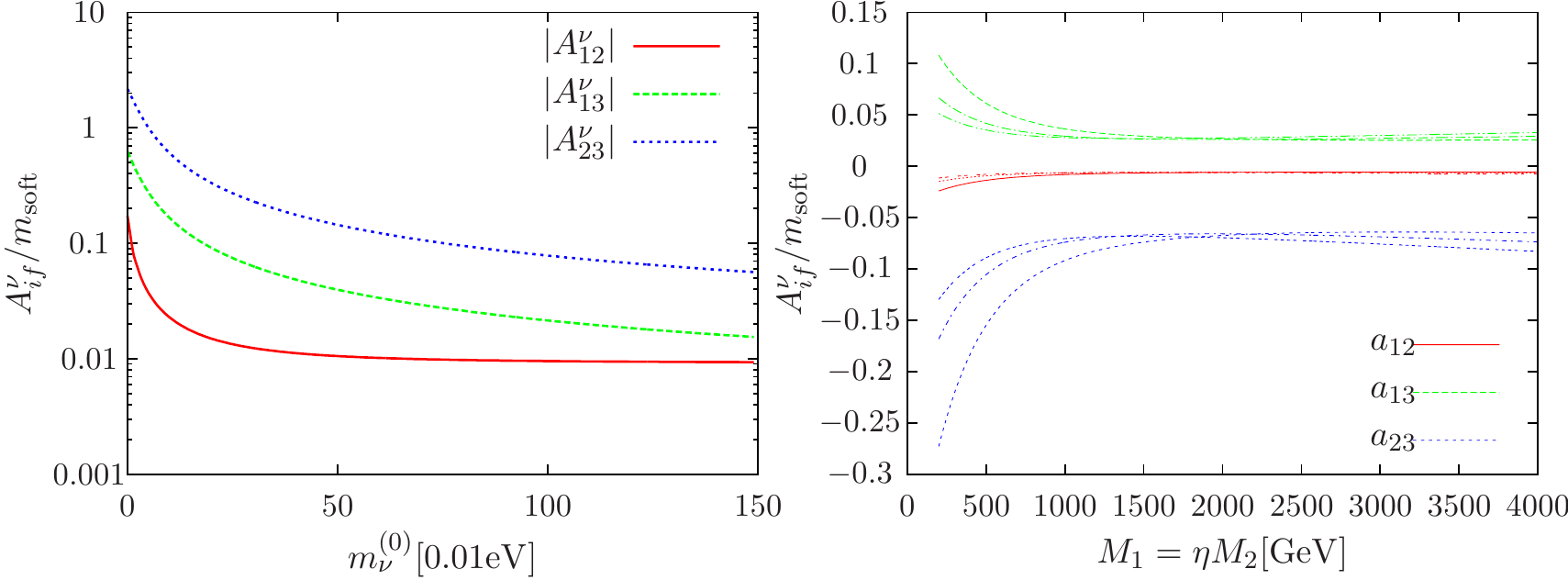}\\
\includegraphics[width=\textwidth]{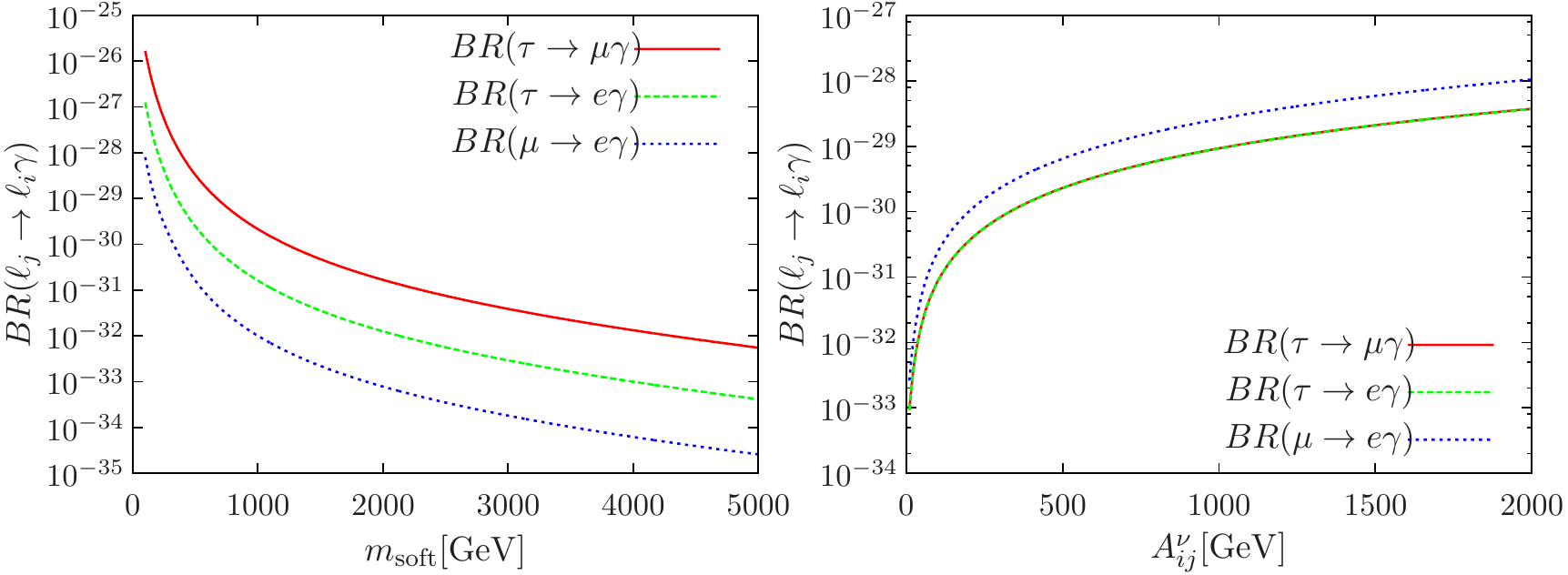}
\caption{We display the results for a sample data point. In the first
  row shows the values of off-diagonal \(A^\nul_{ij}\) in order to
  reproduce the lepton mixing matrix, which is a non-decoupling effect
  as demonstrated by the ratio \(A^\nul_{ij} / m_\text{soft}\). We have
  set a common SUSY mass, \(m_\text{soft} = M_\text{SUSY}\) for scalar
  and gaugino masses. The gaugino masses itself have no influence on the
  result as long as they are not much smaller than \(m_\text{soft}\) as
  shown below. In case of hierarchical neutrinos (\ie with small
  \(m_\nul^{(0)}\)), \(A^\nul_{ij} / m_\text{soft}\) has to be much
  larger to give the same mixing whereas the same ratio may be much
  smaller the more degenerate the neutrino mass spectrum is. In the last
  row, we finally give the contributions to radiative lepton decays
  \(\ell_j \to \ell_i \gamma\) with \(j > i\). Because we only introduce
  off-diagonal in \(A^\nu\) not \(A^e\), the effect on the branching
  ratios is small arising only in the (s)neutrino sector and
  additionally being suppressed with \(M_\Rr\).}\label{fig:result}
\end{figure}

The third (outer right) diagram of Fig.~\ref{fig:seesawreno} obviously
scales as \(\sim A^\nu_{ij} \frac{1}{M_\Rr^2} M^\Rr_{jk} Y^\nul_{kf} /
M_\text{SUSY}\), taking degenerate right-handed neutrinos and (for
simplicity and to see the scaling behavior) \(Y^\nul_{ij} =
y_\nul \delta_{ij}\). The suppression with \(M_\text{SUSY}\) is for
common SUSY masses. So actually, the \(A\)-term contribution to the
neutrino self-energy is
\begin{equation}\label{eq:scaleres}
  \Sigma^{\nul, (A)}_{fi} \sim v_\uq^2 \frac{y_\nul}{M_\Rr}
  \frac{A^\nul_{fi}}{M_\text{SUSY}},
\end{equation}
which is of the same order of magnitude as the tree-level neutrino mass,
\(\sim v_\uq^2/M_\Rr\) and is seen to be suppressed by a typical loop
factor \(g_1^2/(16\pi^2)\). The ratio \(A^\nul_{fi} / M_\text{SUSY}\)
stays the same when \(A^\nul\) and \(M_\text{SUSY}\) are scaled
uniformly. This behavior (using the full analytic expressions with
mixing matrices evaluated numerically) is shown in Fig.~\ref{fig:result}
where we plot several projections of the same dataset. The off-diagonal
values \(A^\nu_{ij}\) are determined such that the renormalized PMNS
matrix of Eq.~\eqref{eq:PMNSreno} with the full expression of \(\Delta
U^\nul_L\) given in Eq.~\eqref{eq:PMNSrenNU} and \(\Delta U^\el_L\)
equal the physical mixing matrix. Furthermore, we have assumed
\(U^{(0)}_{ij} = \delta_{ij}\) to show that it is indeed possible to
generate the full mixing radiatively. The generic soft breaking masses,
especially \(\tilde{m}_\ell^2\) give significant contributions to
charged lepton flavor violation as soon as off-diagonal elements are
considered. Strong constraints on such observables let us arrange
\(\tilde{m}^2_{\ell, e, \nu} = m_\text{soft} \mathds{1}\). A complete
analysis also has to include the proper renormalization of the masses
and constrains also the diagonal terms \(A^\nul_{ii}\).

The mixing matrix renormalization of Fig.~\ref{fig:PMNSreno} relies on
\emph{non-degenerate} neutrino masses. For degenerate masses,
Eq.~\eqref{eq:PMNSrenNU} is ill-defined and nevertheless there is no
need for a renormalization of the mixing
matrix~\cite{Denner:1990yz}. Degenerate neutrino masses, however, allow
for a special treatment and also cause the trivial mixing matrix after
inclusion of threshold corrections to be
non-trivial~\cite{Hollik:2014hya, Hollik:2014tea}.

Non-degenerate masses generically come along with a non-trivial mixing
at tree-level. We have artificially switched it off by a special choice
of the fundamental parameters (\ie both \(M_\Rr\) and \(Y^\nul \sim
\mathds{1}\)). The scale of right-handed neutrinos was chosen high as
suggested by the seesaw mechanism: with couplings \(\lambda_{ij}
\sim\mathcal{O}(1)\) in Eq.~\eqref{eq:WeinbergOp} and the electroweak
scale \(v \sim \mathcal{O}(100\,\GeV)\), we impose \(M_\Rr \sim
\mathcal{O}(10^{13}\,\GeV)\) to have sub-eV neutrinos
\(m_\nul^{(0)} \sim \mathcal{O}(0.1\,\eV)\).

\section{Conclusions}
We have pursued the idea of RFV in the MSSM and applied radiative
techniques to generate lepton mixing in an extension of the MSSM with
right-handed Majorana neutrinos and a seesaw mechanism of type I. Though
the lepton mixing angles are largish compared to the small quark mixing,
the SUSY threshold corrections to the lepton mixing matrix described in
this article are enhanced compared to the corrections to the CKM
matrix. Differently from quarks and charged leptons, the neutrino mass
spectrum is less hierarchic and, depending on the lightest neutrino
mass, can be rather quasi-degenerate. In this case, the large
contributions can be resummed which stabilizes the corrections with
respect to the neutrino mass.\footnote{The resummation has not been
  covered here and was performed after the Corfu Summer Institute 2013
  where the results of this work date back.}

The corrections do not decouple with the SUSY scale and persist if the
SUSY mass spectrum is shifted to higher values. We constrain the
off-diagonal neutrino \(A\)-terms with the requirement that the
renormalized mixing matrix equals the experimentally determined PMNS
matrix. In that description, there is a linear correlation \(U_{ij} \sim
\Sigma^\nul_{ij} \sim A^\nul_{ij}\) between the renormalized mixing
matrix elements, the flavor changing self-energy and the soft SUSY
breaking trilinear coupling. We have shown that this corrections scales
like \(A^\nul_{ij} / M_\text{SUSY}\) resulting in the non-decoupling
behavior with the SUSY scale.

In general, the application of Eqs.~\eqref{eq:PMNSreno}
and~\eqref{eq:PMNSrenNU} to a radiative generation of PMNS elements is
\emph{not} restricted to SUSY theories, however needs a flavor changing
self-energy. Any theory beyond the SM which brings new flavor structures
may lead to such self-energies. The realization within the MSSM (though
extended with right-handed neutrinos) is in line with earlier studies of
RFV and proposes a combined solution of the flavor puzzle together with
SUSY breaking. The origin of flavor may lie in the origin of SUSY
breaking and although the problem is only shifted into a different
sector, the large number of unknown parameters in the general MSSM can
be drastically reduced from radiative flavor physics. Especially, this
formulation does not rely on tree-level flavor symmetries in the
neutrino sector and is not restricted to specific textures in the mass
matrices or Yukawa couplings. What we have not shown here is the
analogous treatment with any non-trivial flavor mixing at the
tree-level. Imposing tribimaximal mixing, \(U^{(0)} = U_\text{TBM}\), we
can equally well generate a non-vanishing 1-3 element and adjust the
other mixing angles to their measured values. This holds for arbitrary
tree-level mixing and emphasizes the importance of flavor non-universal
threshold corrections in the presence of quasi-degenerate neutrinos.

\section*{Acknowledgments}
It is a pleasure to thank the organizers of the Corfu Summer Institute
2013 for their effort to provide an excellent summer school in an
impressive surrounding. The speaker's participation at the summer
institute 2013 as well as the work presented was supported by the
initial research training group GRK 1694 \emph{``Elementarteilchenphysik
  bei h\"ochster Energie und h\"ochster Pr\"azision''} funded by the
Deutsche Forschungsgemeinschaft.


\begin{thebibliography}{99}

\bibitem{Cabibbo:1963yz}
  N.~Cabibbo,
  \emph{Unitary Symmetry and Leptonic Decays},
  Phys.\ Rev.\ Lett.\ {\bf 10} (1963) 531.

\bibitem{Kobayashi:1973fv}
  M.~Kobayashi and T.~Maskawa,
  \emph{CP Violation in the Renormalizable Theory of Weak Interaction},
  Prog.\ Theor.\ Phys.\ {\bf 49} (1973) 652.

\bibitem{Pontecorvo:1957cp}
  B.~Pontecorvo,
  \emph{Mesonium and anti-mesonium},
  Sov.\ Phys.\ JETP {\bf 6} (1957) 429
   [Zh.\ Eksp.\ Teor.\ Fiz.\  {\bf 33} (1957) 549].

\bibitem{Maki:1962mu}
  Z.~Maki, M.~Nakagawa and S.~Sakata,
  \emph{Remarks on the unified model of elementary particles},
  Prog.\ Theor.\ Phys.\ {\bf 28} (1962) 870.

\bibitem{Crivellin:2008mq}
  A.~Crivellin and U.~Nierste,
  \emph{Supersymmetric renormalisation of the CKM matrix and new
    constraints on the squark mass matrices},
  Phys.\ Rev.\ {\bf D79} (2009) 035018,
  \texttt{[arXiv:0810.1613]}.

\bibitem{Weinberg:1972ws}
  S.~Weinberg, \emph{Electromagnetic and weak masses},
  Phys.\ Rev.\ Lett.\ {\bf 29} (1972) 388--392.

\bibitem{Segre:1981nj}
  G.~Segre, \emph{Mass Generation by Radiative Corrections in Minimal SO(10)},
  Phys.\ Lett.\ {\bfseries B103} (1981) 355--358.

\bibitem{Ibanez:1982xg}
  L.~E. Ibanez, \emph{Hierarchical Suppression of Radiative Quark and Lepton Masses
  in Supersymmetric GUTs},
  Phys.\ Lett.\ {\bfseries B117} (1982) 403.

\bibitem{Ibanez:1981nw}
  L.~E. Ibanez, \emph{Radiative Fermion Masses in Grand Unified Theories},
  Nucl.\ Phys.\ {\bfseries B193} (1981) 317.

\bibitem{Lahanas:1982et}
  A.~B.~Lahanas and D.~Wyler,
  \emph{Radiative Fermion Masses and Supersymmetry},
  Phys.\ Lett.\ {\bf B122} (1983) 258.

\bibitem{Masiero:1983ph}
  A.~Masiero, D.~V.~Nanopoulos and K.~Tamvakis,
  \emph{Radiative Fermion Masses in Supersymmetric Theories},
  Phys.\ Lett.\ {\bf B126} (1983) 337.

\bibitem{Banks:1987iu}
  T.~Banks, \emph{Supersymmetry and the Quark Mass Matrix},
  Nucl.\ Phys.\ {\bfseries B303} (1988) 172.

\bibitem{Ma:1988fp}
  E.~Ma,
  \emph{Radiative Quark and Lepton Masses
    Through Soft Supersymmetry Breaking},
  Phys.\ Rev.\ {\bfseries D39} (1989) 1922.

\bibitem{Borzumati:1999sp}
  F.~Borzumati, G.~R. Farrar, N.~Polonsky, and S.~D. Thomas,
  \emph{Soft Yukawa couplings in supersymmetric theories},
  Nucl.\ Phys.\ {\bfseries B555} (1999) 53--115,
  {\ttfamily [arXiv:hep-ph/9902443]}.

\bibitem{Ferrandis:2004ng}
  J.~Ferrandis, \emph{Radiative mass generation and suppression of supersymmetric
  contributions to flavor changing processes},
  Phys.\ Rev.\ {\bfseries D70} (2004) 055002,
  {\ttfamily [arXiv:hep-ph/0404068]}.

\bibitem{Ferrandis:2004ri}
  J.~Ferrandis and N.~Haba,
  \emph{Supersymmetry breaking as the origin of flavor},
  Phys.\ Rev.\ {\bfseries D70} (2004) 055003,
  {\ttfamily [arXiv:hep-ph/0404077]}.

\bibitem{Crivellin:2009pa}
  A.~Crivellin, \emph{CKM Elements from Squark Gluino Loops}, in
  proceedings of the 44th Rencontres de Moriond on Electroweak
  Interactions and Unified Theories, La Thuile 2009,
  {\ttfamily arXiv:0905.3130}.

\bibitem{Crivellin:2011sj}
  A.~Crivellin, L.~Hofer, U.~Nierste, and D.~Scherer,
  \emph{Phenomenological consequences of radiative flavor violation in the MSSM},
  Phys.\ Rev.\ {\bfseries D84} (2011) 035030,
  {\ttfamily [arXiv:1105.2818]}.

\bibitem{Crivellin:2011fb}
  A.~Crivellin, L.~Hofer, and U.~Nierste,
  \emph{The MSSM with a Softly Broken \(\U(2)^3\) Flavor Symmetry},
  in proceedings of EPS-HEP 2011,
  \pos{PoS(EPS-HEP2011)145} (2011),
  {\ttfamily [arXiv:1111.0246]}.

\bibitem{Altmannshofer:2014qha}
  W.~Altmannshofer, C.~Frugiuele, and R.~Harnik,
  \emph{Fermion Hierarchy from Sfermion Anarchy},
  JHEP {\bfseries 1412} (2014) 180,
  {\ttfamily [arXiv:1409.2522]}.

\bibitem{Denner:1990yz}
  A.~Denner and T.~Sack,
  \emph{Renormalization of the Quark Mixing Matrix},
  Nucl.\ Phys.\ {\bf B347} (1990) 203.

\bibitem{Agashe:2014kda}
  {\bfseries Particle Data Group} , K.~Olive {\em et~al.},
  \emph{Review of Particle Physics},
  Chin.\ Phys.\ {\bfseries C38} (2014) 090001.

\bibitem{Weinberg:1979sa}
  S.~Weinberg,
  \emph{Baryon and Lepton Nonconserving Processes},
  Phys.\ Rev.\ Lett.\ {\bfseries 43} (1979) 1566--1570.

\bibitem{Minkowski:1977sc}
  P.~Minkowski,
  \emph{\(\mu \to e \gamma\) at a Rate of One Out of 1-Billion Muon
    Decays?},
  Phys.\ Lett.\ {\bfseries B67} (1977) 421.

\bibitem{Mohapatra:1979ia}
  R.~N. Mohapatra and G.~Senjanovic,
  \emph{Neutrino Mass and Spontaneous Parity Violation},
  Phys.\ Rev.\ Lett.\ {\bfseries 44} (1980) 912.

\bibitem{Magg:1980ut}
  M.~Magg and C.~Wetterich,
  \emph{Neutrino Mass Problem and Gauge Hierarchy},
  Phys.\ Lett.\ {\bfseries B94} (1980) 61.

\bibitem{Schechter:1980gr}
  J.~Schechter and J.~W.~F.~Valle,
  \emph{Neutrino Masses in \(\SU(2)\times\U(1)\) Theories},
  Phys.\ Rev.\ {\bf D22} (1980) 2227.

\bibitem{Schechter:1981cv}
  J.~Schechter and J.~W.~F.~Valle,
  \emph{Neutrino Decay and Spontaneous Violation of Lepton Number},
  Phys.\ Rev.\ {\bf D25} (1982) 774.

\bibitem{Foot:1988aq}
  R.~Foot, H.~Lew, X.~He, and G.~C. Joshi,
  \emph{Seesaw Neutrino Masses Induced by a Triplet of Leptons},
  Z.\ Phys.\ {\bfseries C44} (1989) 441.

\bibitem{Gonzalez-Garcia:2014bfa}
  M.~C.~Gonzalez-Garcia, M.~Maltoni and T.~Schwetz,
  \emph{Updated fit to three neutrino mixing: status of leptonic CP violation},
  JHEP {\bf 1411} (2014) 052,
  \texttt{[arXiv:1409.5439]}.

\bibitem{Hollik:2014lka}
  W.~G.~Hollik,
  \emph{(Quasi-)Degeneration, Quantum Corrections and Neutrino Mixing}, in
  proceedings of the 42th ITEP Winter School, Moscow 2014,
  \texttt{arXiv:1411.2946}.

\bibitem{Hollik:2014jda}
  W.~G.~Hollik and U.~J.~Salda\~na Salazar,
  \emph{The double mass hierarchy pattern: simultaneously understanding
    quark and lepton mixing},
  Nucl.\ Phys.\ {\bf B892} (2015) 364,
  \texttt{[arXiv:1411.3549]}.

\bibitem{Dedes:2007ef}
  A.~Dedes, H.~E.~Haber and J.~Rosiek,
  \emph{Seesaw mechanism in the sneutrino sector and its consequences},
  JHEP {\bf 0711} (2007) 059,
  \texttt{[arXiv:0707.3718]}.

\bibitem{Hollik:2014hya}
  W.~G.~Hollik,
  \emph{Radiative generation of neutrino mixing: degenerate masses and
    threshold corrections},
  Phys.\ Rev.\ {\bf D91} (2015) 033001,
  \texttt{[arXiv:1412.4585]}.

\bibitem{Hollik:2014tea}
  W.~G.~Hollik,
  \emph{Lifting degenerate neutrino masses, threshold corrections and
    maximal mixing}, in proceedings of the Young Scientists Workshop
  ``Flavorful Ways to New Physics'', Freudenstadt 2014,
  \pos{PoS(FWNP)018} (2014)
  \texttt{[arXiv:1412.5117]}.
\end{thebibliography}
\end{document}